# 5G Radio Access above 6 GHz

Mehrdad Shariat[1], David M. Gutierrez-Estevez[1], Arnesh Vijay[2], Krystian Safjan[2], Patrik Rugeland[3], Icaro da Silva[3], Javier Lorca[4], Joerg Widmer[5], Maria Fresia[6], Yilin Li[7], Isabelle Siaud[8]

[1]Samsung Electronics R&D Institute UK, [2]Nokia Bell Labs, [3]Ericsson AB, [4]Telefonica I+D, [5]IMDEA Networks, [6]Intel Deutschland GmbH, [7]Huawei Technologies Dusseldorf GmbH, [8]Orange

*Abstract*— Designing and developing a millimetre-wave (mmWave) based mobile Radio Access Technology (RAT) in the 6-100 GHz frequency range is a fundamental component in the standardization of the new 5G radio interface, recently kicked off by 3GPP. Such component, herein called the new mmWave RAT, will not only enable extreme mobile broadband (eMBB) services, but also support UHD/3D streaming, offer immersive applications and ultra-responsive cloud services to provide an outstanding Quality of Experience (QoE) to the mobile users. The main objective of this paper is to develop the network architectural elements and functions that will enable tight integration of mmWave technology into the overall 5G radio access network (RAN). A broad range of topics addressing mobile architecture and network functionalities will be covered— starting with the architectural facets of network slicing, multi-connectivity and cells clustering, to more functional elements of initial access, mobility, radio resource management (RRM) and self-backhauling. The intention of the concepts presented here is to lay foundation for future studies towards the first commercial implementation of the mmWave RAT above 6 GHz.

## I. INTRODUCTION

There has been a significant growth in the volume of mobile data traffic in recent years and this trend is expected to continue, reaching at least 6500 Peta Bytes/month by 2017, corresponding to a 7-fold increase over 2013 [1]. This is due to the proliferation of smart phones and other mobile devices that support a wide range of newly deployed broadband applications and services. In 5G, eMBB services [2] will play an important role, however, with more challenging requirements such as very high peak data rates (in the order of 10 Gbps) in certain eMBB scenarios [3] and, at least, 50 Mbps everywhere including urban, suburban and rural areas served by the mmWave [3]. It is also envisioned that 5G should support a new range of services, such as ultra-reliable machine applications, which may require very low latencies (in the order of 1 ms) and ultra-reliable connectivity to serve applications such as traffic safety, infrastructure protection or emerging industrial Internet applications. In order to meet these requirements, a new RAT is also being studied by 3GPP [4], where it is assumed that it should operate at frequencies up to 100 GHz. In order to reach these goals, there is a consensus in the mobile industry [2], [5] that mmWave technology must be an essential component of that new 5G RAN, not only as a capacity enhancer for low frequency deployments in macro and/or localized hotspots, but to enable robust connectivity for users connected only to the mmWave RAT without any support from a RAT operating below 6 GHz. This could either be mmWave networks deployed independently from low frequency RATs or more likely co-deployed with only partial overlapping coverage (e.g. indoor deployment) where the low frequency support cannot be obtained in the entire area. In mmWave bands, the high carrier frequency and large bandwidth, in addition to the maintenance of fragile links in high frequencies for mobile users, influence the system concept and technological components. To support a diverse set of services, scenarios and frequency bands, the 5G systems will comprise a combination of technologies, requiring the mmWave technology to integrate in an inter-operable manner.

To enable the integration of mmWave within the new 5G RAN, it will be important to facilitate the integration at the lowest layer possible in the protocol stack, instead of relying on higher layers for inter-RAT communication. This is important in order to enable fast resource management, User Plane aggregation and to exploit multi-connectivity solutions.

In this paper, we develop initial concepts on network architectural elements and functions that will enable tight integration of mmWave technology with other 5G system components. The outlined solutions provide robust inter-operability between mmWave carriers and other 5G frequency carriers (including also upcoming carrier(s) by the evolution of LTE-A), thereby providing an 'edge-less' user experience. Furthermore, it is anticipated that mmWave technology can enable other aspects in future 5G systems. For this, our work investigates possibilities for mmWave-based wireless backhaul dimensioning along the radio access part. Hence, flexible and scalable network deployments can be enabled to reduce network costs and to increase their adaptability to dynamic conditions.

The remainder of this paper is organized as follows: Section II presents requirements and challenges of mmWave networks, covering 5G use cases (UCs) and spectrum requirements, channel characterization, radio interface constraints and limitations and further extends into discussing the RAN design challenges. Section III presents the proposed RAN architecture, covering various topics of network slicing, multi-connectivity, multi-layer and multi-RAT management, mmWave access point (AP) clustering, network dimensioning and backhauling. In Section IV, important RAN functionalities are presented, covering the subject of initial access, mobility management, Medium Access Control (MAC) / RRM aspects, and self-backhauling. Finally, conclusions are drawn in Section V.

## II. REQUIREMENTS AND CHALLENGES

### A. 5G Use Cases and Spectrum Requirements

Several 5G UCs have been identified in different regulatory bodies, standardization fora and research programmes (e.g., 5G PPP projects) where there are ongoing efforts to harmonize them. Some of the most important UCs which may benefit from the mmWave technologies, either on the access, backhaul, or both, are outlined in Fig. 1 [6]. In the wide spectrum range from 6 to 100 GHz different bands are suitable for different use cases. Use cases such as tactile internet or robotic control will be preferably served in lower part of 6-100 GHz spectrum whereas high end of this range will be used for use cases from eMBB family - such as cloud service or immersive 5G early experience.

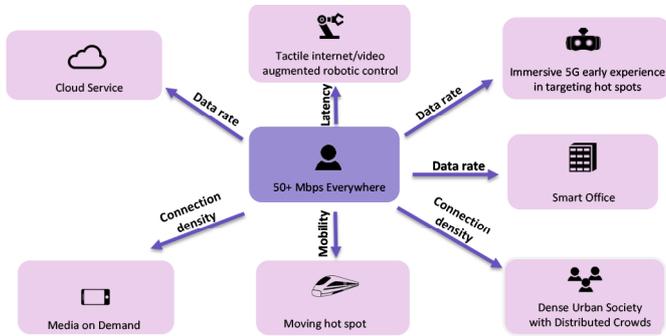

Fig.1 Key mmWave use cases [6]

These UCs set stringent requirements on the networks, e.g. support for peak data rates of 10 Gbps, reduced latency in the order of 5 ms (end-to-end) for interactive applications and a highly improved and consistent QoE to users, regardless of the UE location or mobility.

The support of a wide contiguous bandwidth is a key factor in most of the aforementioned UCs. Therefore, multiple spectrum band options need to be explored in the so-called mmWave spectrum in the 6-100 GHz range. Each band has its own unique features including radio propagation characteristics, contiguity, as well as co-primary allocations. In particular, some key factors including capacity, coverage, mobility, and device complexity can be explored to characterize each candidate band. As an example, mmWave in the low GHz range (6-30 GHz) can generally provide better coverage at a lower device complexity (due to the low directivity gain required, thus simplifying the antenna design). However, limitation on the availability of contiguous vacant spectrum may be a limiting factor on the provision of capacity. On the other hand, in the mid to high GHz range (above 30 GHz), a wider vacant spectrum is available―but higher directivity and beam-forming gain is required to compensate for the higher path loss, increasing the device complexity and also impacting the mobility management and/or system access procedures due to fine beam tracking requirements.

### B. Channel Characterization

The propagation characteristics in the mmWave bands significantly differ from the characteristics below 6 GHz, which heavily impacts RAN design. Apart from the fact that the path loss scales with frequency, radio propagation moves more towards the behaviour of optical signals with lower diffraction, and increasingly relies upon line of sight (LOS) path or strong reflections rather than diffuse components. At lower base station heights and higher frequencies, the direct path or strong multi-path components are more likely to be blocked―with the effect that the signal power can drop significantly, rapidly and unpredictably. Blockage can be induced by trees and street furniture, traffic or people. Furthermore, at higher frequencies, even small differences in angle and path length between the LOS and reflected paths can cause strong spatial fading up to a breakpoint distance.

Great efforts have been made and are currently in progress to develop comprehensive advanced propagation and channel models within the entire frequency range from 6 to 100 GHz for various 5G deployment scenarios [7]. The proposed initial channel model is in line with the 3GPP-3D model, and an open source platform for simulations has been made available in form of the QuaDRiGa implementation [8], covering the frequency range from 10 to 80 GHz. Also additional requirements and extensions to the existing models have been identified as being potentially crucial in order to accurately support the mmWave frequency range [7]. In this direction, extensive measurement campaigns have been conducted and are in full progress involving advanced mmWave channel sounder setups with multi-frequency (up to four bands in parallel) and ultra-wideband capabilities (up to 4 GHz bandwidth). Evaluation results of the first campaigns have been described for the scenarios of urban micro-cellular street canyon, open square, outdoor to indoor, office, shopping mall and airport.

### C. Radio interface constraints and limitations

The use of mmWave technologies imposes specific challenges to the radio interface (RI), compared to sub-6 GHz frequencies. Link budget constraints resulting from higher isotropic free-space loss needs to be compensated by higher antenna gains via directional transmission. Fortunately, such high frequencies allow smaller antenna elements, enabling the deployment of the large antenna arrays required for high-gain beamforming. In terms of transceiver architectures, the standard solution will likely support both analog and digital architectures. However, in practice, the directional transmission may need to rely on hybrid transceiver architectures (with both analog and digital processing) to save on hardware cost and power consumption. The mmWave RI also requires high resolution and fast Analog to Digital (ADC) and Digital to Analog (DAC) converters to provide the high data rates envisioned in the GHz range. Directional transmissions can change the effective channel characteristics, e.g. interference characteristics, leading to different requirements and design principles for RI development. Furthermore, critical Radio Frequency (RF) / hardware impairments that increase with carrier frequency must be taken into account, such as phase noise, In-phase / Quadrature (I/Q)-imbalance, sampling jitter, sampling frequency offset, carrier frequency offset and power amplifier (PA) non-linearity. Such impairments can lead to increased Error Vector Magnitude (EVM) and reduced spectral efficiency. In addition, asymmetric antenna and RF configurations in uplink (UL) and downlink (DL) also affect RI design, considering the fact that UL coverage will be constrained by much lower TX power at the user equipment (UE).

*D. RAN Design Challenges*

Several key challenges are involved in the design of a RAN which includes a mmWave RAT. Some challenges are related to the flexibility and configurability required in future networks to support a versatile set of services and business models by expanding emerging concepts like network slicing to the RAN level. Other challenges arise from specific characteristics in the mmWave bands (e.g. directional and outage prone transmissions), requiring a redesign of some RAN architectural aspects (multi-connectivity, multi-layer and multi-RAT management, AP clustering, and network dimensioning) as well as functional aspects (initial access, mobility management, RRM and self-backhauling).

In the following sections, we will explore in more details how these challenges can be addressed and what are the enabling technologies from architectural and functional perspective.

## III. RAN ARCHITECTURE

The 5G RAN architecture design has been studied in several research projects [3] and is currently studied for standardization by 3GPP [4]. The architecture is assumed to use LTE as a baseline. In this section, we explore some of the modifications and extensions to LTE that are required to address the 5G use cases.

*A. Impact of Network slicing on 5G RAN design*

Network slicing is an important part of the vision for the overall 5G architecture [2] that addresses the deployment of multiple logical networks as independent business operations on a common physical infrastructure. The ultimate goal would be to provide "network slices on an as-a service basis" and to meet the wide range of use cases that the 2020 time frame will demand for different industries [10].

A slice is seen by a slice customer as an independent network. However, in contrast to deploying an independent network infrastructure, each slice will be realized together with other slices on a common infrastructure (also referred to as "virtual network"), including assets such as hardware, software or radio resources. As the propagation conditions of the mmWave are expected to vary frequently as described in Section II.B, it will be important to avoid assigning dedicated radio resources to specific slices in order to seamlessly release unused radio resources and to share available resources with other slices. In this way, the infrastructure and asset utilization will be much more cost and energy-efficient.

The concept of network slicing has initially been proposed for the 5G core network (CN) [10]. However, the notion has been recently applied end-to-end and it is possible that RAN may need specific functionality to support multiple slices or even partitioning of resources for different network slices [11]. To implement the functionality required for network slicing, there are several baseline assumptions that can be made on the architecture and RAN design.

- Sharing: Most of the network resources in the RAN architecture are shared by multiple slices by default;
- Differentiation: The 5G RAN enables mechanisms for per-slice traffic differentiation;
- Visibility: The 5G RAN should have the necessary visibility to apply slice differentiation;
- Protection: The 5G RAN should provide protection mechanisms to minimize inter-slice effects;
- Management: The 5G RAN should provide the support for slice-specific network management.

*B. Multi-connectivity*

Multi-connectivity is a key component to fulfil 5G requirements on data-rate, latency, reliability, and availability. In particular, the channel characteristics and inherent fragility of the mmWave links will require redundant and complementary coverage to ensure reliable connectivity. The term multi-connectivity can accommodate a broad range of techniques all of which provide a given UE with radio resources from two or more radio links.

Inter-frequency multi-connectivity refers to the case where a UE is connected to two interfaces on different carrier frequencies, either from a single RAT or from two different RATs (including bands where the evolution of LTE-A might be deployed). In contrast, intra-frequency multi-connectivity refers to transmissions on the same frequency, e.g. Coordinated Multipoint (CoMP) schemes.

Depending on the deployment and use case requirements, different options may be selected based on a trade-off between throughput, reliability or latency, on the one hand, and signalling overhead and complexity on the other hand. The options for inter-frequency multi-connectivity can either be based on carrier aggregation (CA) specified by 3GPP from LTE Rel-10 [12] or based on dual connectivity (DC) specified by 3GPP from LTE Rel-12 [13].

For CA the traffic is split at the MAC layer which requires tight time synchronization between the carriers. This restricts the use to either carriers from a single node or to deployments where the nodes have inter-site connections with very low latencies (e.g. via fibre or between collocated nodes) and where the numerologies are aligned (if the nodes belong to different RATs).

For DC, the traffic is split at or above the Packet Data Convergence Protocol (PDCP) layer where one of the nodes is the Master eNB (MeNB) controlling the data split and the other is the Secondary eNB (SeNB) which relays the traffic. This allows for a much more relaxed time synchronization, which simplifies inter-site and inter-RAT DC. It is envisioned to enhance the DC (i.e. eDC) by allowing traffic to split in both MeNB and SeNB without relocating MeNB. Various considered service flows (SFs), i.e. Master Cell Group (MCG) SF and Secondary Cell Group (SCG) SF as well as SF Splits at Master 5G-NB (M5G-NB) and Secondary 5G-NB (S5G-NB) are shown in Fig. 2. Additionally, it will be possible to split the traffic between more than two eNBs, simultaneously.

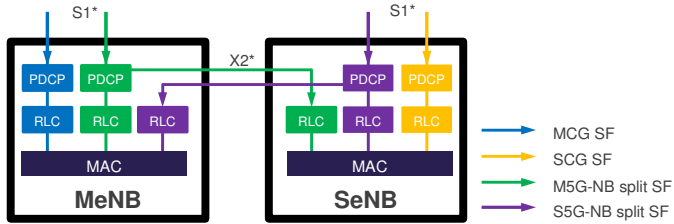

Fig. 2 Radio protocol architecture for 5G enhanced dual connectivity

Multi-connectivity could include only the user plane, only the control plane or both depending on the requirements and the capacity, reliability, and latency of each link. Furthermore, it is important to assess which technique is preferable in different scenarios.

*C. Multi-layer and Multi-RAT management*

To achieve multi-layer and multi-RAT integration, the logical functions required for switching between different RATs can be abstracted at different layers. The proposed generic model may consider several layers for integration.

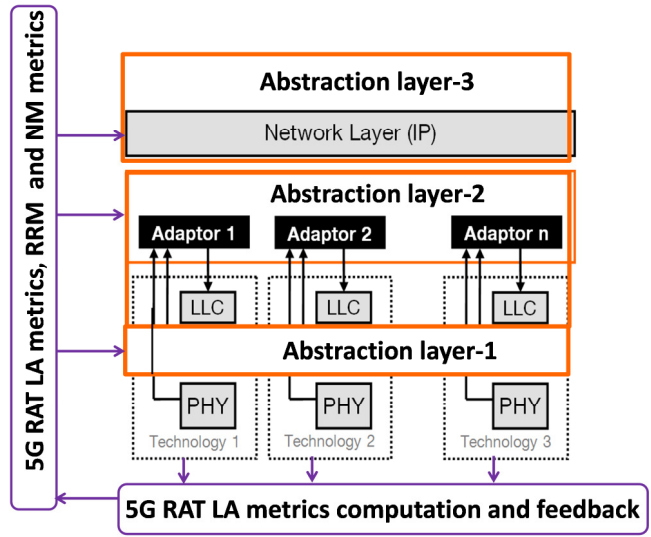

Fig. 3 Multi-layer multi-RAT management architecture

Fig. 3 illustrates the multi-layer multi-RAT concept where three layers are envisioned to perform RAT switching in a unified structure. This multi-layer multi-RAT architecture can also be denoted as "multi-layer abstraction" linked to NGMN 5G concepts.

The abstraction layer-1 performs transmission mode switching via exploiting MAC mechanisms in existing protocols. This is achieved by integrating new link adaptation metrics (using the same communication channel) to forward the decision [14].

The abstraction layer-2 considers a L2.5 layer managing several air interfaces when control cannot be shared between them. This solution is an extension of the I-MAC concept [15] by integrating multiple criteria in the multi-RAT management.

The abstraction layer-3 exploits IP network protocols to switch from one interface, slice or network, to another. This level of abstraction can also be utilized for Control / User (C/U) plane splitting between different RATs (e.g. WiFi vs. cellular) in order to perform offloading in small cells.

*D. mmWave AP clustering*

mmWave APs will likely be deployed very densely in many scenarios. mmWave links can break easily and may need to be re-established quickly. Therefore, clustering of mmWave APs is crucial to deal with this inherent network dynamics. A cluster is defined as a group of APs in the vicinity of a UE, capable of serving that UE. The APs included in the cluster can be configured by the network and subsequently reconfigured when the UE moves. To coordinate the mobility within a cluster, one of the APs can be designated as the cluster head (CH) which is connected to the CN through S1* interface (as an enhanced version of current S1 interface) [3]. The CH is also connected to all other APs in the cluster. Depending on the topology of the network and the capacity and latency of the backhaul links, the network position and role of the CH may vary.

The network structure for a cluster with sufficiently well-dimensioned backhaul can be made quite similar to that of CoMP. All the intelligence is located in a central node (the CH) which is responsible for all control and user plane protocol handling, including how the mmWave beams are tracked at different APs. The CH decides which APs serve the UE or which APs stay in stand-by mode.

For non-ideal backhaul or wireless (self-) backhauling, the CH location depends on the topology of the cluster. Besides the coordination on intra-cluster mmWave beam switching, the CH handles the majority of data sent to and received from the UE. In order for the CH to coordinate the inter-AP switching in a fast and efficient manner, it is assumed that there is only one hop between the CH and the APs in the cluster.

*E. Network dimensioning and backhauling*

An important facet which can influence the RAN of 5G is the subject of backhauling. An improved version of wired and wireless backhauling or a combination of the two can be deployed in future mobile systems. The specific requirements from the RAN, together with the planned deployment scenarios, determine the design requirements for the selected backhaul type. Early 5G scenario requires means for incremental deployment, as initially the density of 5G base stations with dedicated backhauling is limited. A useful technique which can possibly be beneficial in future systems is self-backhauling.

Self-backhauling refers to a set of solutions where coverage/capacity extension is applied to a network/infrastructure by means of APs utilizing the same radio resources used by the UEs to access the network, as shown in Fig. 4. Self-backhauling provides an efficient way to combat infrastructure constraints especially in dense network deployment where access to fibre may be limited to only some APs. However, over time as the fixed infrastructure will become more available, the self-backhauling will gradually evolve.

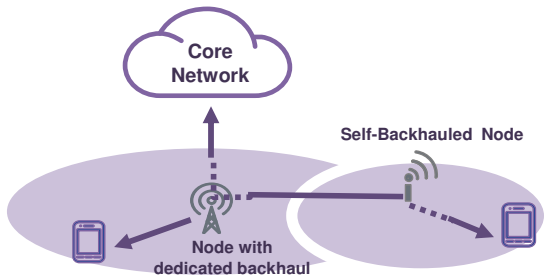

Fig. 4 Concept of self-backhauling

When specifically considering self-backhauling in mmWave frequencies, they can bring about great benefits for future radio systems, but in-parallel can encompass a unique set of challenges. The problem lies in the inherent directionality property of mmWave bands, which can result in frequent link blockage and subsequent spotty coverage due to street furniture and objects present in the surrounding environment. Focusing on this issue, some ways to tackle can be:

- From a physical layer perspective, developing advanced channel estimation and beamforming algorithms can assist in exploiting non-LOS paths that make use of reflections from the surrounding environment.
- When considering the MAC and network layer, tighter integration and coordination will be necessary between the adjacent mmWave APs to form dynamic beam clusters to circumvent obstacles in line-of-sight paths. Here, it becomes imperative to support efficient backhauling solutions between the adjacent mmWave access points and those of other technologies. For this, wireless mesh backhaul topology can be quite advantages when compared to initial colossal cost incurred when deploying wired (e.g. fiber) solutions.
- Lastly, the multi-hop backhauling condition can be an interesting solution for moving hotspots, which can offer dynamic paths towards the core network via intermediate mmWave APs in its vicinity.

A key challenge is that each network hop adds to the latency whilst reducing the overall capacity. Therefore, in order to design a multi-hop system without compromising the end-to-end latency requirements, techniques such as fast scheduling cycles and rapid rerouting are crucial. More details on self-backhauling functionalities will be covered in the next section.

IV. RAN FUNCTIONALITIES

In this section we explore RAN functionalities to enable the mmWave RAT to operate at frequencies above 6 GHz.

A. Initial access

The initial access to the mmWave RAT is a critical element and needs to be designed for both standalone and non-standalone deployments as the coverage of the mmWave RAT may only be partially overlapping with the coverage of RATs below 6 GHz, (e.g. with indoor deployments).

The initial access can be separated into three tasks which need to be possible to perform both in the standalone and non-standalone scenario:

- Downlink timing/frequency synchronization,
- System information acquisition, and
- Uplink timing synchronization (if required).

The performance of the applied initial access scheme directly impacts the user experience. One of the performance-critical challenges is to achieve beam alignment within a short time interval during the initial access phase, for example, by exploiting available a-priori information on the preferred transmission direction at both ends of the link. If the UE is already connected to a low frequency RAT (e.g. LTE-A) the initial access could be more efficient if some of the functionalities were performed by the low frequency RAT compared cases where there is no low-frequency coverage.

To enable coverage detection, the mmWave AP is assumed to transmit some synchronization signals (SSs) at particular time-frequency resources with a certain periodicity in order for the UE to obtain the cell identity (ID). To achieve fast mmWave small cell detection, the low frequency RAT can further transmit a signal to the UE, containing information about the frequency and cell IDs of the mmWave small cells within its coverage area. With this signaling, the UE does not need to perform an exhaustive search over the whole small cell ID space, but it only tries to detect the signaled cell IDs. As a consequence, the UE power consumption for downlink synchronization can be significantly reduced.

The coverage of the system information determines the coverage of the cell. Some of the system information components are changing fast on the basis of one or several mmWave RAT frames but other system information parts may vary relatively slow. For this reason, it can be energy efficient to convey some of the slowly varying system information by exploiting an existing low frequency RAT.

Uplink (UL) synchronization needs to be achieved prior to any UL packet transmission to ensure that all the co-scheduled UEs' UL signals arrive at the eNB around the same time within the cyclic prefix (CP) duration. The radio resources for the preamble transmission are typically signaled in the system information and, as mentioned above, such system information can be signaled by the low frequency RAT if the UE is anyway monitoring those frequencies.

*B. Mobility management*

Efficient mobility management in the mmWave RAT is required for a seamless service experience for users on the move. As the mmWave RAT supports multi-connectivity (between different mmWave nodes or inter-RAT between mmWave RAT and LTE-A), the mobility management should support the required coordination between the nodes even when the transport network to which they are connected is capacity-limited or adds up latency.

As outlined earlier, mmWave AP clustering is instrumental to support mobility in a mmWave RAT. Intra-AP beam switching from one beam to another within a cluster can be supported by UE measurement feedback and direct communication between UE and AP on the beam switch.

In case of inter-AP beam switching, the report will be forwarded to the CH from the current serving AP. The CH will request the target AP for beam switching. If positive feedback is received from the target AP, the UE will be eventually informed (via CH and serving AP) to switch its beam and will be served via target AP after the switch. In case of inter-RAT handover, the maximum data rate supported by the technology and expected rate fluctuations (as a result of a handover) may be shared as cross-layer information with application level services to avoid frequent rate variations, impacting the end user QoE.

*C. MAC and RRM*

Challenging path loss conditions are a key driver for performing some advanced joint RRM techniques over multiple cells in mmWave bands. Beamforming can lead to bursty inter-cell interference in ultra-dense scenarios where strong multipath components from reflectors can impact users camping in adjacent cells. Some coordination between cells can be effective in coping with this strong and unpredictable interference. RRM coordination can rely on the user device as an anchor node to avoid costly exchange of inter-cell information, with the twofold objective of i) detecting any interfering beams from neighbor cells and ii) reporting this condition to the serving cell. Beams can be conveniently labeled upon transmission with a unique identifier containing e.g. the intended UE to be served. A victim UE can acquire this information and together with the interfering cell ID, send an "interference report" to the serving cell to inform of this condition. Interference reports from multiple UEs can then serve as a basis for further coordination of resources among the cells involved, either in the temporal, frequency or spatial (beam) domain.

Hybrid Automatic Repeat-Request (HARQ) operation at MAC level can also be challenging in mmWave systems. Given the large bandwidths foreseen to be available in mmWave bands, fulfilling the HARQ round trip time (RTT) may be challenging in centralized deployments, leading to significant delay constraints in the transport network. Moreover, increasing the HARQ RTT in the uplink may impact the device complexity as transmission buffer length will increase, leading to device issues. One alternative may be to decouple uplink HARQ from Forward Error Correction (FEC) decoding, by closing the HARQ loop at the antenna whilst performing full FEC decoding of the packets in a central node at a later time. This approach can help to relax the transport network delay requirements without increasing the device complexity

*D. Self-backhauling aspects*

While considering self-backhaul design requirements, various parameters such as latency, reliability, availability and average/peak-UL/DL data rates must be taken into account. In addition, other attributes such as availability of frequency bands, spectrum regulation and mobility access requirements will also have to be considered.

The dynamics and self-autonomy of self-backhauling solutions can gradually evolve into Software Defined Networking (SDN)-based solutions, where one logical controller is supposed to monitor topology changes, node-to-node radio channel status and all the traffic needs in a real-time manner.

In this case, in one envisioned scenario backhaul networking for densely deployed small cells could be characterized by a ringed-tree topology [16] with multiple backhaul links per node and different levels of backhaul links (where vertical links would have higher priorities in route selections than horizontal ones). An example of a ringed-tree backhaul networking is illustrated in Fig. 5.

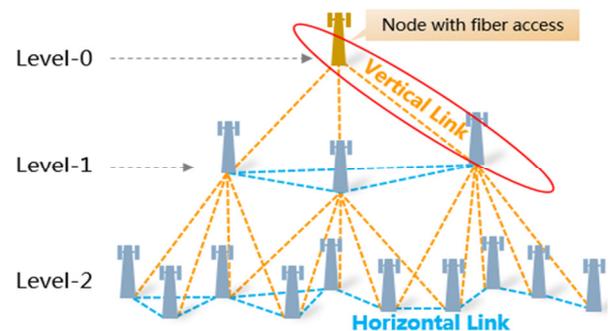

Fig. 5 An example of ringed-tree backhauling

Moreover, dynamic RRM decisions will be made by the controller in terms of how much radio resources to be allocated per link at each node. The network can also enable multi-route connection, allowing coordination and cooperation among network elements to be performed via management provided by the controller (in order to achieve network-level optimization).

In the context of radio resource management, multiplexing of time and frequency resources between backhaul and access links attracts the most attention, where the two links are operating on the same frequency band in a dynamic TDD system. In this kind of scenario, one relevant issue is to determine efficient resource allocation mechanisms to split radio resources between backhaul and access links, ensuring that there are enough backhaul capacities for all traffic from access links in order to maximize the resource utilization of the network. Another relevant issue is that when both backhaul and access links operate in the same spectrum band, the optimum distribution of non-conflicting resources between the two links needs to be determined. In addition to the time resource management between backhaul and access links under in-band relaying in a dynamic TDD system, multiplexing of spatial resources that exploits directional transmission with extensive beamforming has been envisioned to be a promising candidate to overcome the unfavorable path loss in high frequency bands.

## V. CONCLUSIONS

The so-called mmWave band (6-100 GHz) is envisioned to play a key role in enabling highly advanced services and applications such as eMBB or UHD/3D streaming for future 5G systems. In this paper, we reviewed key requirements and challenges for mmWave networks followed by a description of key RAN architectural components and functionalities of the mmWave-based RAT that will enable tight integration of mmWave technology into the overall 5G system. On the RAN architecture side, we covered key aspects of network slicing, multi-connectivity, multi-layer and multi-RAT management, access point clustering, and network dimensioning and backhauling. On the RAN functionality side, we presented contributions on initial access, mobility management, MAC/RRM, and self-backhauling aspects. All these technology components will lay the foundations for future studies towards a commercially feasible mmWave RAT within future 5G systems.

## ACKNOWLEDGMENT

The research leading to these results received funding from the European Commission H2020 programme under grant agreement n°671650 (5G PPP mmMAGIC project).